# Workflow for investigating thermodynamic, structural and energy properties of condensed polymer systems from Molecular Dynamics

**James P. Andrews and Estela Blaisten-Barojas***

*Center for Simulation and Modeling and Department of Computational and Data Sciences, George Mason University, 4400 University Drive, Fairfax, Virginia 22030, USA*

*Correspondence: blaisten@gmu.edu

March 8, 2022

ABSTRACT

Soft matter materials and polymers are widely used in the controlled delivery of drugs. Simulation and modeling provide insight at the atomic scale enabling a level of control unavailable to experiments. We present a workflow protocol for modeling, simulating, and analyzing structural and thermodynamic response properties of poly-lactic-co-glycolic acid (PLGA), a well-studied and FDA approved material. We concatenate a battery of molecular dynamics, computational chemistry, highly parallel scripting, and analysis tools for generating properties of bulk polymers in the condensed phase. We provide the workflow leading to the glass transition temperature, enthalpy, density, isobaric heat capacity, thermal expansion coefficient, isothermal compressibility, bulk modulus, sonic velocity, cohesive energy, and solubility parameters. Calculated properties agree very well with experiments, when available. This methodology has been extended to a variety of polymer types and environments.

## 1. Introduction

Currently, polymeric nanoparticles with hollow core as those formed from poly lactic-*co*-glycolic acid (PLGA) constitute a vast class of soft matter nanostructures of exceptional technological significance for drug delivery and tissue engineering applications.[1,2] PLGA is a biocompatible and FDA-approved biodegradable copolymer that has been used for the controlled delivery of antibiotics and macromolecules such as DNA and RNA. The benefits of such a delivery system are multiple, including applications in vaccines. Drug dosages and speed of delivery have been studied empirically by tuning the synthesis for specific physical properties of the condensed polymer phases. However, there is no systematic study at the atomic level of PLGA glassy solid, rubbery/leather region and liquid phases.

In this work we describe a computational protocol for the scientific analysis of PLGA form the perspective of all-atom simulations of the PLGA polymer system. The workflow concatenates a collection of molecular dynamics packages, computational chemistry software, and numerous custom codes and databases tools developed in a high performance computing environment specifically for investigating the thermodynamic, structural and energy properties of PLGA. This workflow is applied for a system of PLGA(50:50) with polymer chains containing 222 monomers.

## 2. Protocol Methodology and Results

PLGA is a co-polymer composed of two kinds of monomers, lactic acid (L) and glycolic acid (G) of which lactic acid has a pair of stereo-isomers, the L- and D-lactic acid. Initial polymer chains were constructed using a custom computer code able of creating arbitrary length polymer chains with an arbitrary sequence of monomer types. The unoptimized chemical structures of L-L, D-L, and G monomers were from PubChem. A polymeric matrix of PLGA(50:50) was built containing a certain number of chains, each chain formed by an equal proportion of L and G monomers randomly sequenced along the chain. In these polymer chains, half of the L monomers were of type L-L and the other half of type D-L. For this study ten chains of 24 monomers each were created as described, completing the workflow step identified as *Monomer sequence* in Fig. 1. The ten 24-mer chain models were energetically optimized at the density functional theory level using the package Gaussian09.[3] These optimized structures and corresponding atomic charges are ported to Antechamber (AmberTools package)[4] to generate parameters for the GAFF force field. Our custom script allows for selection between two schema of atomic charges: the restrained electrostatic potential (RESP) or the Bond Charge Correction (BCC) method[5]. This process completes the other two workflow steps identified as *atomic charge refinement* and *force field* in Fig. 1.

Longer chains of 222-mers were coalesced by stitching together combinations of the 22 central monomers of the different 24-mer chain models using Leap (AmberTools package)[4]. A total of twelve polymer chains of molecular weight 14,459 u were generated. These stitched chains were relaxed and allowed to acquire random coil conformations in vacuum. The four steps of the workflow are implemented in an embarrassingly parallel workload manner. Thereafter, these 222-mer chains populated a computational box for modeling a PLGA(50:50) system with 20,016 atoms, completing the workflow step *Structural optimization* in Fig. 1. The latter constituted the actual initial system configuration. Thereafter, in step *NPT annealing* of Fig. 1, the GOMACS package[6] was used for equilibrating the system with molecular dynamics (MD) within the constant number of particles N, constant pressure P and constant temperature T approach NPT at a high temperature of T=500 K and P=101.325 kPa. The MD parameters were: periodic boundary conditions, time step of 1 fs, cutoff radius of 1.4 nm, velocity rescaling temperature and Berendsen pressure couplings, PME (particle-mesh-Ewald) for long range electrostatics, and the system was set to run for 20 ns. Aided with a battery of scripts for the job scheduler Slurm, thirty MD-NPT sequential runs were concatenated for annealing the system in a descending ladder-wise fashion from 500 K to 200 K. Each simulated temperature followed from the ending configuration of the previous temperature simulation.

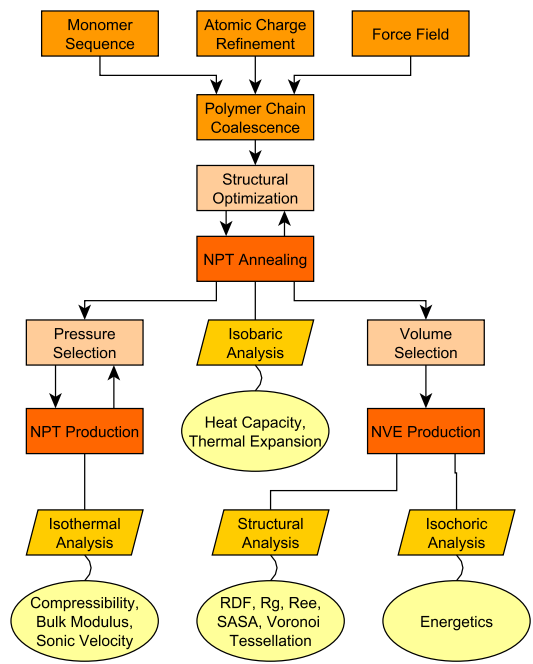

Figure 1. Workflow of tasks leading to the all-atom inspection of the PLGA(50:50) polymer in its condensed phases.

The fundamental outcome of the stepwise temperature annealing was the generation of the caloric curve of enthalpy as a function of temperature (Fig. 2a), its fluctuations (Fig. 2b), its Legendre transform (Fig. 2c), and the system volume as a function of temperature (Fig. 2d), which constituted the workflow task *Isobaric analysis* of Fig. 1. A custom optimization routine for identifying the inflection point between linear regions of the enthalpy, enthalpy variance, and volume led to a predicted glass transition temperature ($T_g$) and its bracketing temperature region between $T_{g\alpha}$ and $T_{g\beta}$ along which the polymer matrix transitioned from a viscosity-dominated system to a mechanically elastic material. The final step of the workflow central block was *Specific heat, thermal expansion*, which included the calculation of both properties from the change of enthalpy and of volume with respect to temperature, respectively.

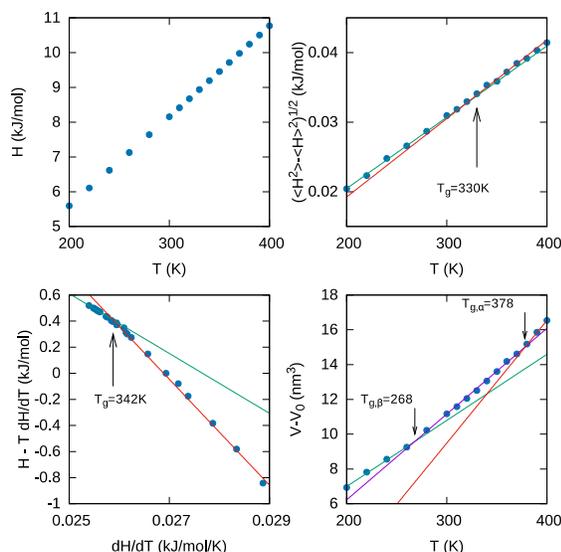

Figure 2. (a) PLGA enthalpy H(T) as function of temperature. (b) The enthalpy standard deviation as a function of temperature. (c) The Legendre transform of H(T). (d) System volume V related to the volume at low temperature $V_o$. The glass transition temperature $T_g$ is depicted with arrows.

From here on, the workflow tasks split into two branches emanating from the step *NPT annealing* of Fig. 1. For the righthand side branch, the strategy was to use a round-robin scheduler in which as soon as one of the MD-NPT ladder-descending temperature runs finished, a new MD-NVE simulation at the equilibrated volume was initiated and run for 20 ns. The latter involved steps *Volume selection* and *NVE Production* depicted in Fig. 1. Next, two parallel tasks in the workflow, (i) *Isochoric analysis* and (ii) *Structural analysis*, led in (i) to energy derived properties obtained with custom built codes (cohesive energy, cohesive energy density, Hildebrand parameter) and in (ii) to radial distribution functions (rdf) and solvent-accessible surface area (SASA) calculated with GROMACS accessible tools. In addition, custom analysis codes and scripts were written for polymer chain radius of gyration (Fig. 3), moments of inertia, configuration tensor, chain volumes, among other chain-based properties. Meanwhile, the lefthand side branch of the workflow relates to a process designed for obtaining the polymer system behavior at a select set of pressures with the final goal of determining the isothermal compressibility of PLGA(50:50). This task entailed several MD-NPT runs at six pressures in the range 500-50,000 kPa for each of four temperatures, 200 K, 300 K, 400 K, 500 K, involving the workflow tasks *Pressure selection*, *NPT Production* and *Isothermal analysis* of Fig. 1. The latter task enabled the calculation of the thermal compressibility coefficient and the two related properties, bulk modulus and sonic velocity. This was the last task of the workflow.

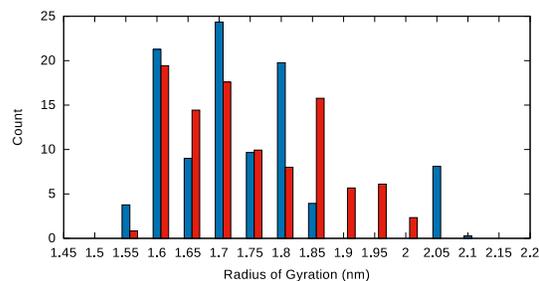

Figure 3. Distribution of the radius of gyration collected from MD-NVE runs of 20 ns at T=300 K (blue) and T=500 K (red). The histogram collects $R_g$ values for the twelve chains of length 222-mers along a 20 ns MD-NVE run.

## 3. Discussion

In a nutshell, Table 1 gives the most relevant properties of the PLGA(50:50) condensed phases obtained along the multiple tasks performed by our workflow. Calculation of the glass transition is not trivial. The calculated glass transition, Tg was 330 K from the fluctuations of the enthalpy and 342 K from the Legendre transform of H(T) illustrated in Figs. 2b and 2c, respectively. This Tg is higher than experimental values in the range 318-328 K[11]. However, the temperature bracket around $T_g$ of $262 \pm 8$ to $365 \pm 11$ that was obtained and shown in Fig. 2c is consistent with measurements[12]. Most experiments are done for PLGA systems of molecular weight larger than the one used here. Comparing with literature published values, our results are in very good agreement, reinforcing the validity of our predicted results, which await experimental confirmation.

Table 1. Calculated PLGA(50:50) properties for a system with twelve chains with 222-mers each.

| | $T=300$ | $T=400$ | Experiment |
|---|---|---|---|
| Density $\rho$ (kg/m$^3$) | $1319.1 \pm 0.7$ | $1288 \pm 1$ | $1340$ [a] |
| Heat capacity $C_p$ (J/kg/K) | 2977 | | |
| $C_p$ (J/kg/K) | 2921 | | |
| Thermal expantion $\alpha$ ($10^{-4}$K$^{-1}$) | $2.1 \pm 0.2$ | | |
| Thermal compressibility $\kappa_T$ (GPa$^{-1}$) | $0.14 \pm 0.01$ | $0.21 \pm 0.01$ | |
| Bulk modulus $B$ (GPa) | $7.0 \pm 0.3$ | $4.7 \pm 0.3$ | $\approx 4$ [b] |
| Sonic velocity $s$ (m/s) | $2318 \pm 39$ | $1915 \pm 56$ | $2326 - 2450$ [c,d] |
| Cohesive energy/mer $E_{coh/mer}$ (kJ/mol) | $19.13 \pm 0.06$ | $18.23 \pm 0.08$ | |
| Hildebrand parameter $\delta_h$ (MPa$^{1/2}$) | $19.69 \pm 0.03$ | $18.98 \pm 0.04$ | |
| (a) [7]; (b) [8]; (c) [9]; (d) [10] | | | |

## 4. Conclusion

In this presentation we have described the numerous steps involved in modeling, simulating, and analyzing structural and thermodynamic response properties of poly-lactic-co-glycolic acid (PLGA). Our protocol included exploration of a number of properties for the polymer under study, including the glass transition temperature, enthalpy, density, isobaric heat capacity, thermal expansion coefficient, isothermal compressibility, bulk modulus, sonic velocity, cohesive energy, and solubility parameter of PLGA. This approach is portable for the study of other polymers or complex materials.

Our findings are important in the design of nanostructures and nanoparticles as well as in the control of polymer folding when devising scaffoldings for tissue engineering. Part of this workflow was already successfully applied for analyzing the effects of solvents on the structure of solvated PLGA[5]. Currently, we have extended the workflow use for the analysis of PLGA(50:50) with a variety of molecular weights[13] and in nanocomposite systems[14].

**Acknowledgments**

We acknowledge partial support from the Commonwealth of Virginia 4-VA grant *Scalable Molecular Dynamics*. JA is thankful for the Provost Office presidential scholarship support. All computations were done in the high performance computing clusters of the Office for Research Computing, George Mason University.